# Strong coupling expansion of chiral models

Massimo Campostrini, Paolo Rossi, and Ettore Vicari [a] [*]

[a]Dipartimento di Fisica dell'Università di Pisa, Italy.

A general precedure is outlined for an algorithmic implementation of the strong coupling expansion of lattice chiral models on arbitrary lattices. A symbolic character expansion in terms of connected values of group integrals on skeleton diagrams may be obtained by a fully computerized approach.

## 1. Introduction.

The main motivations for a strong-coupling approach to asymptotically free matrix-valued field theories (like 2-d principal chiral models) are the following:

(i) Search for a scaling region, within the convergence radius of the strong coupling series, where the continuum results may be extracted by appropriate resummation. Numerical evidence favors this possibility [1].

(ii) Extension of the scaling region, essentially by adopting the "energy" scheme [3] in the definition of the coupling directly in the strong coupling series and studying convergence and scaling properties. Numerical evidence indicates asymptotic scaling within the strong coupling domain [1,2].

(iii) Study of the relevance of the planar limit $N \to \infty$, which in turn leads to an analysis of the double-scaling limit properties of many-matrix systems, in particular a determination of the location and nature of the criticality and evaluation of the relevant critical exponents. Numerical evidence points at a second order transition at $\beta \simeq 0.306$ for the 2-d chiral models on the square lattice [2].

## 2. Strong coupling expansion.

The strong coupling expansion is best performed by adopting the character expansion of the action [4,5]

$$e^{2N\beta \mathrm{Re} \mathrm{Tr} U} = e^{N^2 F(\beta)} \sum_{(r)} d_{(r)} z_{(r)} \chi_{(r)}(U), \quad (1)$$

[*]This work was partially supported by MURST, and E.C. contract CHRX-CT92-0051.

where $F(\beta)$ is the free-energy of the single-link model, $\chi_{(r)}$ are the characters and $d_{(r)}$ the dimensions of the irreducible representations of $U(N)$ and $z_{(r)}$ are the character coefficients and are naturally organized into a strong-coupling series. Notable simplifications occur in the large-$N$ limit of the strong coupling expansion: we must however notice that strong coupling and large-$N$ limit are not strictly commuting, which implies that some care is needed in the extrapolations, that are however supported by numerical evidence.

In order to sketch the logic of an algorithmic implementation of the strong coupling expansion we must introduce a few definitions. Lattice functional integrals are reduced by the character expansion to sums of group integrals on nontrivial assignments of representations on the lattice links. Assignments can be grouped into "oriented configurations" characterized only by the number of "quarks" $n_+$ and "antiquarks" $n_-$ running on each lattice link. Configurations ($n_+ + n_- = n$ constant) can be turned into skeleton diagrams, where each (abstract) link is characterized by $n$ and the length of the associated lattice path. Configurations sharing the same skeleton have the same value of the group integral. We therefore define the "geometrical factor" as the number of configurations on each skeleton, and the "group-theoretical factor" as the value of the group integral for each skeleton.

The geometrical factor can be evaluated by a fully computerized procedure:

- generate all nonbacktracking random walks with fixed length, i.e. fixed order in strong coupling;

- keep one and only one walk for each different configuration;
- identify the skeletons of configurations and compare them to each other implementing permutation symmetries;
- collect all configurations with equivalent skeleton.

The group-theoretical factor requires computing group integrals, and is in principle a completely solved problem, but in practice it requires generating algorithmically the Clebsch-Gordan coefficients for the decomposition of products of arbitrary irreducible representations of $U(N)$. Some progress in the evaluation of group-theoretical factors is obtained by eliminating bubbles in skeletons diagrams and reducing the problem to evaluating only superskeletons group integrals (a much smaller number of objects): superskeletons are diagrams where no multiple connections between sites are allowed; they are obtained by use of the simpler decomposition of products of characters.

By known results of group integration [6], all integrals relevant to rather high orders can be explicitly computed. By exploiting geometric properties of oriented configurations, we may evaluate the connected contribution of each skeleton to the expectation value of a physical quantity. Connected group-theoretical factors are called "potentials": thay are the quantities admitting a finite large-$N$ limit. Wide families of potentials, related to the same superskeleton, can be computed symbolically in terms of the $z_{(r)}$ and numerically in terms of $N$ and $\beta$. The symbolic character expansion in terms of potentials is obtained by a fully computerized process.

### 3. The fundamental Green's function.

A few important considerations concern Green's functions

$$G(x) = \frac{1}{N} \langle \text{Tr} \left[ U(x) U(0)^\dagger \right] \rangle ,$$
$$\widetilde{G}(p) = \sum_x e^{ipx} G(x) . \quad (2)$$

Truncated strong coupling expansions admitting a reinterpretation as summations over paths can be see as originated by generalized Gaussian models with non-nearest interactions. At $q^{th}$ order in strong coupling expansion the inverse propagator admit the expansion

$$\widetilde{G}(p)^{-1} = A_0(\beta) + A_1(\beta)\hat{p}^2$$
$$+ \sum_{u=2}^{q/3} \sum_{\substack{s=0 \\ u-s=\text{even}}}^{u} A_{u,s}(\beta) \hat{p}^{2s} \left[ \left(\hat{p}^2\right)^2 - \hat{p}^4 \right]^{(u-s)/2} \quad (3)$$

where the number of independent coefficients $A_{u,s} \sim \beta^{3u}$ is equal to the number of independent effective couplings. Strong coupling series for $A_{u,s}$ are an efficient way of storing all the strong coupling information concerning $G(x)$.

Physical quantities may be extracted directly from $A_{u,s}$:
- magnetic susceptibility: $\chi = 1/A_0$.
- second-moment correlation length: $\langle x^2 \rangle_G = A_1/A_0$.
- wavefunction renormalization: $Z_G = 1/A_1$.
- True mass gap $\mu_s$ by inversion of

$$x = 2(1 - \text{ch}\mu_s) , \quad (4)$$

where $x$ is the perturbative solution of

$$A_0 + A_1 x + \sum_{u=2}^{\infty} A_{u,u} x^u = 0 . \quad (5)$$

Thermodynamical quantities like energy and specific heat may also easily be extracted.

### REFERENCES

1. P. Rossi and E. Vicari, Phys. Rev. **D**, 49 (1994) 6072.
2. M. Campostrini, P. Rossi and E. Vicari, "Asymptotic scaling from strong coupling", IFUP-TH 36/94 (1994).
3. G. Parisi, in Proceedings of the XXth Conference on High Energy Physics, Madison, Wisconsin, 1980.
4. F. Green and S. Samuel, Nucl. Phys. B190 (1981) 113.
5. J.M. Drouffe and J.B. Zuber, Phys. Rep. 102 (1983) 1.
6. B. de Wit and G. 'tHooft, Phys.Lett. 69B (1977) 61.